\documentclass[12pt]{article}
\usepackage{graphics,graphpap,color}
\usepackage{graphicx}

\newcommand\bmat{\left( \begin{array}{cc}}
\newcommand\emat{\end{array}\right)}

\def\msbar{\ifmmode{\overline{\rm MS}} \else{$\overline{\rm MS}$} \fi}
\def\drbar{\ifmmode{\overline{\rm DR}} \else{$\overline{\rm DR}$} \fi}

\def\ti              {\tilde}

\def\a               {\alpha}
\def\b               {\beta}
\def\d               {\delta}
\def\D               {\Delta}

\def\g               {\gamma}
\def\G               {\Gamma}

\def\t               {\theta}

\def\sf              {{\ti f}}

\def\sfi             {{\ti f_i^{}}}
\def\sfL             {{\ti f_L^{}}}
\def\sfR             {{\ti f_R^{}}}

\def\st              {{\ti t}}
\def\sb              {{\ti b}}
\def\stau            {{\ti \tau}}

\newcommand{\msf}[1]   {m_{\ti f_{#1}}}

\def\tW              {\t_{\scriptscriptstyle W}}

\def\non             {\nonumber}

\renewcommand\d{\delta}

\begin{document}

\pagestyle{empty} \vspace*{-1cm} 

\begin{flushright}
  HEPHY-PUB 766/02 \\
  hep-ph/0210354
\end{flushright}

\vspace*{2cm} 

\begin{center}
{\Large\bf\boldmath
   Full electroweak one--loop corrections \\ to {\boldmath{ $A^0 
   \rightarrow \tilde{f}_i \,\ {\bar{\!\!\tilde{f}}}_{\!\!j}$}}}\\[5mm]

\vspace{10mm}

\underline{C.~Weber}, H.~Eberl and W.~Majerotto\\[5mm]

\vspace{6mm}
{\it Institut f\"ur Hochenergiephysik der \"Osterreichischen 
Akademie der Wissenschaften, A--1050 Vienna, Austria} 

\vspace{6mm}
{\it Contribution to SUSY02, 10th International Conference on 
Supersymmetry and Unification of Fundamental Interactions, 17--23 
June 2002, DESY Hamburg, Germany.} 
\end{center}

\vspace{20mm}

\begin{abstract}
We discuss the full electroweak one--loop corrections to the decay 
of the pseudoscalar Higgs boson $A^0$ into two sfermions within 
the Minimal Supersymmetric Standard Model. In particular, we 
consider the sfermions of the third generation, $\st_i$, $\sb_i$ 
and $\stau_i$, including the left--right mixing. The electroweak 
corrections can go up to $\sim15\%$ and can therefore not be 
neglected. 
\end{abstract}

\vfill
\newpage
\pagestyle{plain} \setcounter{page}{2}

\section{Introduction}
The Minimal Supersymmetric Standard Model (MSSM) \cite{MSSM} 
requires five physical Higgs bosons: two neutral CP--even ($h^0$ 
and $H^0$), one heavy neutral CP--odd ($A^0$), and two charged 
ones ($H^\pm$) \cite{GunionHaber1, GunionHaber2}. The existence of 
heavy neutral Higgs bosons would provide a conclusive evidence of 
physics beyond the SM. Therefore, searching for Higgs bosons is 
one of the main goals of future collider projects like TEVATRON, 
LHC or an $e^+ e^-$ Linear Collider. 

In this talk, we consider the decay of the CP--odd Higgs boson 
$A^0$ into two sfermions, $A^0 \rightarrow \tilde{f}_i \,\ 
{\bar{\!\!\tilde{f}}}_{\!\!j}$. The decays into sfermions can be 
the dominant decay modes of Higgs bosons in a large parameter 
region if the sfermions are relatively light \cite{Bartl1, 
Bartl2}. In particular, third generation sfermions $\st_i$, 
$\sb_i$ and $\stau_i$ can be much lighter than the other sfermions 
due to their large Yukawa couplings and their large left--right 
mixing. We have calculated the full electroweak corrections in the 
on--shell scheme and have implemented the SUSY--QCD corrections 
from \cite{SUSY-QCD}. We will show that the electroweak 
corrections are significant and need to be included. 

At tree--level the Higgs sector depends on two parameters, for 
instance $m_{A^0}$ and $\tan\beta$. $m_{A^0}$ is the mass of the 
pseudoscalar Higgs boson $A^0$, and $\tan\b = \frac{v_2}{v_1}$ is 
the ratio of the vacuum expectation values of the two neutral 
Higgs doublet states \cite{GunionHaber1, GunionHaber2}. In the 
chargino and neutralino systems there are the higgsino mass 
parameter $\mu$, the $U(1)$ and $SU(2)$ gaugino mass parameters 
$M'$ and $M$, respectively. We assume that the gluino mass 
$m_{\tilde g}$ is related to $M$ by $m_{\tilde g} = 
(\a_s(m_{\tilde g})/\a_{2})\sin^2\theta_W M$. 

\vspace{2mm}
\section{Tree--level Width}
The decay width for $A^0 \rightarrow \tilde{f}_i \,\ 
{\bar{\!\!\tilde{f}}}_{\!\!j}$ depends on the left--right mixing. 
This mixing is described by the sfermion mass matrix in the 
left--right basis $(\sfL, \sfR)$, and in the mass basis $(\sf_1, 
\sf_2)$, $\sf = \st, \sb$ or $\stau$,
\begin{eqnarray}
  {\cal M}_{\sf}^{\,2} \,=\,
   \left( 
     \begin{array}{cc} 
       m_{\sf_L}^{\,2} & a_f\, m_f
       \\[2mm]
       a_f\,m_f & m_{\sf_R}^{\,2}
     \end{array}
   \right)
  = \left( R^\sf \right)^\dag
   \left( 
     \begin{array}{cc} 
       m_{\sf_1}^{\,2} & 0
       \\[2mm]
       0 & m_{\sf_2}^{\,2}
     \end{array}
   \right) R^\sf \,,
\end{eqnarray}
where $R^\sf_{i\a}$ is a 2 x 2 rotation matrix with rotation angle
$\theta_{\!\sf}$, which relates the mass eigenstates $\sf_i$, $i = 
1, 2$, $(m_{\sf_1} < m_{\sf_2})$ to the gauge eigenstates 
$\sf_\a$, $\a = L, R$, by $\sf_i = R^\sf_{i\a} \sf_\a$ and
\begin{eqnarray}
  m_{\sf_L}^{\,2} &=& M_{\{\ti Q\!,\,\ti L \}}^2
       + (I^{3L}_f \!-\! e_{f}^{}\sin^2\!\tW)\cos2\b\,m_Z^{\,2}
       + m_{f}^2\,, \\[2mm]
  m_{\sf_R}^{\,2} &=& M_{\{\ti U\!,\,\ti D\!,\,\ti E \}}^2
       - e_{f}\sin^2\!\tW \cos2\b\,m_Z^{\,2}
       + m_f^2\,, \\[2mm]
  a_f &=& A_f - \mu \,(\tan\b)^{-2 I^{3L}_f} \,.
\end{eqnarray}
$M_{\ti Q}$, $M_{\ti L}$, $M_{\ti U}$, $M_{\ti D}$ and $M_{\ti E}$ 
are soft SUSY breaking masses, $A_f$ is the trilinear scalar 
coupling parameter, $I^{3L}_f$ and $e_f$ are the third component 
of the weak isospin and the electric charge of the sfermion $f$, 
and $\theta_W$ is the Weinberg angle.\\ The mass eigenvalues and 
the mixing angle in terms of primary parameters are
\begin{eqnarray}
  \msf{1,2}^2  
    &=& \frac{1}{2} \left(
    \msf{L}^2 + \msf{R}^2 \mp
    \sqrt{(\msf{L}^2 \!-\! \msf{R}^2)^2 + 4 a_f^2 m_f^2}\,\right)
\\
  \cos\t_{\sf}
    &=& \frac{-a_f\,m_f}
    {\sqrt{(\msf{L}^2 \!-\! \msf{1}^2)^2 + a_f^2 m_f^2}}
  \hspace{2cm} (0\leq \t_{\sf} < \pi) \,,
\end{eqnarray}
and the trilinear breaking parameter $A_f$ can be written as
\begin{eqnarray}\label{mfAf}
m_f A_f  = \frac{1}{2}
\left(m_{\sf_1}^2-m_{\sf_2}^2 \right) \sin 2\theta_\sf \,+\, m_f
\, \mu \,(\tan\b)^{-2 I^{3L}_f} \,.
\end{eqnarray}
At tree--level the decay width of $A^0 \rightarrow \tilde{f}_i \,\ 
{\bar{\!\!\tilde{f}}}_{\!\!j}$ is given by
\begin{eqnarray}
\G^{\rm tree}(A^0 \rightarrow \tilde{f}_i \,\ 
{\bar{\!\!\tilde{f}}}_{\!\!j}) &=& \frac{N_C^f\, \kappa 
(m_{A^0}^2, m^2_{\sf_i}, m^2_{\sf_j})}{16 \,\pi\, m^3_{A^0}}\ 
|G_{ij}^{\sf}|^2 
\end{eqnarray}
with $\kappa (x, y, z) = \sqrt{(x-y-z)^2 - 4 y z}$ and the colour 
factor $N_C^f = 3$ for squarks and $N_C^f = 1$ for sleptons 
respectively. The Higgs--Sfermion--Sfermion couplings for the 
pseudoscalar Higgs boson $A^0$ are given by 
\begin{eqnarray}
G_{ij}^{\sf} &=& \frac{i}{\sqrt2}\, h_f \Bigg( \hspace{-2pt} A_f 
\Bigg\{ \hspace{-3pt}
\begin{array}{rr}\cos\beta \\ \sin\beta
\end{array}
\hspace{-3pt} \Bigg\} + \mu \Bigg\{ \hspace{-5pt}
\begin{array}{rr}\sin\beta \\
\cos\beta \end{array} \hspace{-3pt} \Bigg\} \Bigg)
\hspace{-3pt}\left( \hspace{-5pt} \begin{array}{rr} 0 ~& 1 \\ -1 
~& 0 
\end{array} \right)_{\!\!ij}
\end{eqnarray}
for {\textrm{\scriptsize{$\left\{\hspace{-4pt}\begin{array}{cc}
{\textrm{\footnotesize{up}}}\\
{\textrm{\footnotesize{down}}}
\end{array}\hspace{-3pt}\right\}$}}}--type
sfermions respectively. $h_f$ denotes the Yukawa couplings \\ $h_t 
= g\, m_t/(\sqrt{2} m_W \sin\b), h_b = g\, m_b/(\sqrt{2} m_W 
\cos\b)$ and $h_\tau = g\, m_\tau/(\sqrt{2} m_W \cos\b)$ for top, 
bottom and tau, respectively. 

\vspace{2mm}
\section{Electroweak Corrected Decay Width}

\begin{figure}[th]
\begin{picture}(160,215)(0,0)
     \put(0,-2){\mbox{\resizebox{16cm}{!}
     {\includegraphics{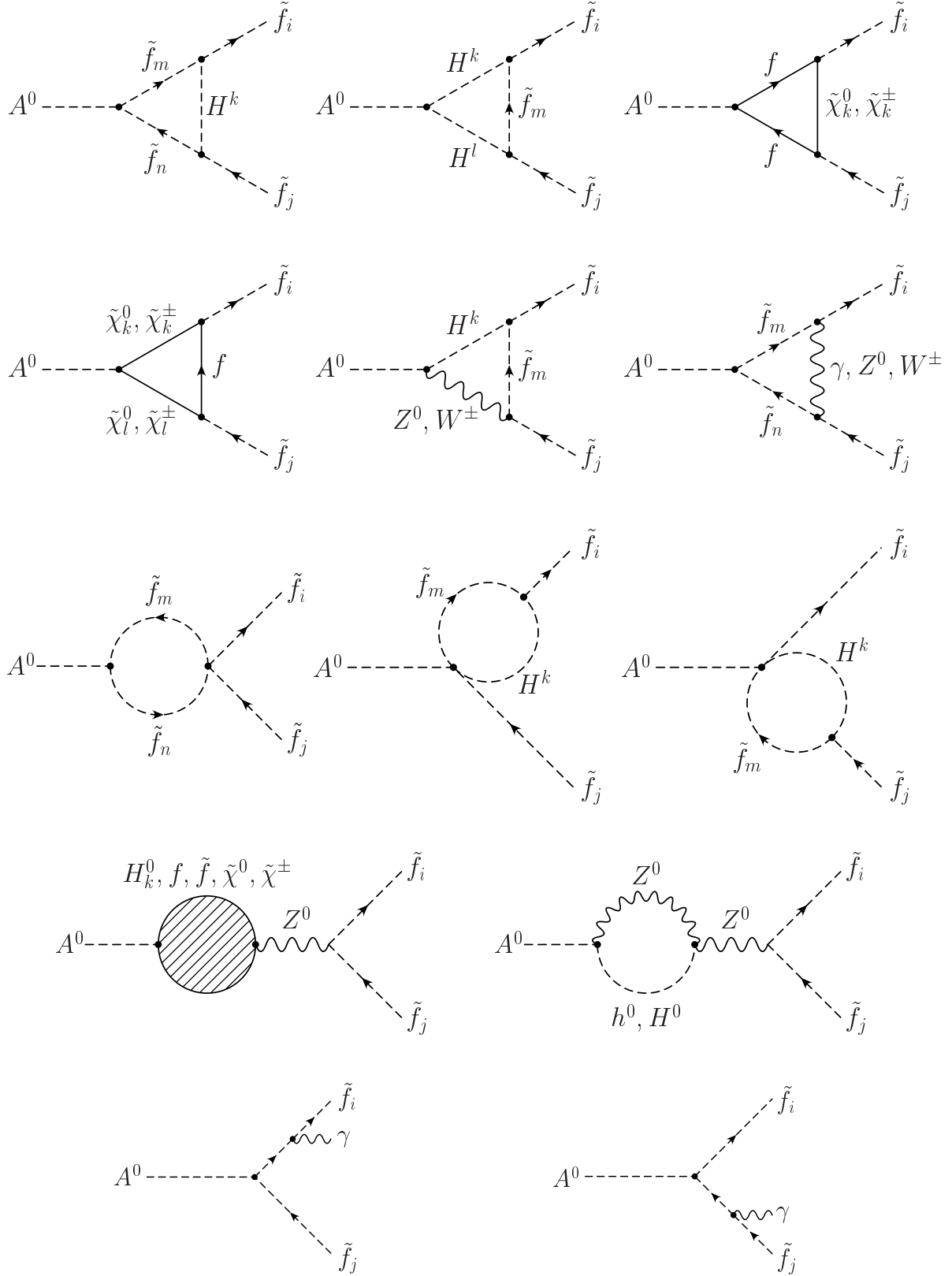}}}}
\end{picture}
\caption{Vertex and photon emission diagrams relevant to the 
calculation of the virtual electroweak corrections to the decay 
width $A^0 \rightarrow \tilde{f}_i \,\ 
{\bar{\!\!\tilde{f}}}_{\!\!j}$. \label{vertex-graphs}}
\end{figure}

The one--loop corrected (renormalized) amplitude $G_{ij}^{\sf\, 
\rm ren}$ can be expressed as 
\begin{eqnarray}
G_{ij}^{\sf\, \rm ren} &=& G_{ij}^{\sf} ~+~ \d G_{ij}^{\sf} ~=~
G_{ij}^{\sf} ~+~ \d G_{ij}^{\sf (v)} ~+~ \d G_{ij}^{\sf (w)} ~+~ 
\d G_{ij}^{\sf (c)} \,, 
\end{eqnarray}
where $G_{ij}^{\sf}$ denotes the tree--level 
$A^0$--$\sfi$--$\sf_j$ coupling in terms of the on--shell 
parameters, $\d G_{ij}^{\sf (v)}$ and $\d G_{ij}^{\sf (w)}$ are 
the vertex and wave--function corrections, respectively. Here we 
only show the diagrams of the vertex graphs 
(Fig.~\ref{vertex-graphs}).
Note that in addition to the one--particle irreducible vertex 
graphs also one--loop induced reducible graphs with $A^0$--$Z^0$ 
mixing have to be considered. Since all parameters in the coupling 
$G_{ij}^{\sf}$ have to be renormalized, the counter term 
correction reads 
\begin{eqnarray} \d G_{ij}^{\sf (c)} &=& 
\frac{\d h_f}{h_f}\, G_{ij}^{\sf} + \frac{i}{\sqrt2}\, h_f \ \d \! 
\left( \hspace{-2pt} {A_f} \left\{ \hspace{-5pt} 
\begin{array}{rr}{\cos\beta} \\ {\sin\beta}
\end{array} \hspace{-5pt} \right\} + \mu \left\{ \hspace{-5pt}
\begin{array}{rr}{\sin\beta} \\
{\cos\beta} \end{array} \hspace{-5pt} \right\} \right) \,.
\end{eqnarray}
\clearpage\noindent %
The Yukawa coupling counter term can be decomposed into 
corrections to the electroweak coupling $g$, the masses of the 
fermion $f$ and the gauge boson $W$ and the mixing angle $\b$,
\begin{eqnarray}
   \frac{\d h_f}{h_f} &=& \frac{\d g}{g} \,+\, \frac{\d m_f}{m_f} 
   \,-\, \frac{\d m_W}{m_W} \,+\, {\left\{ \hspace{-5pt}
   \begin{array}{r}{-\cos^2\beta} \\ {\sin^2\beta}
   \end{array} \hspace{-3pt} \right\}}
   \frac{\d\tan\beta}{\tan\beta} \,.
\end{eqnarray}
For the trilinear coupling we get with eq.~(\ref{mfAf})
\begin{eqnarray}
   \frac{\d A_f}{A_f} &=& \frac{\d (m_f A_f)}{m_f A_f} -
   \frac{\d m_f}{m_f} \,,
\\[2mm] \non
   \d(m_f A_f) &=& \d \!\left(\! m_f \mu\, {\left\{ \hspace{-5pt}
   \begin{array}{rr}{\cot\beta} \\ {\tan\beta}
   \end{array} \hspace{-5pt} \right\}} \!\right) \,+\, \frac{1}{2}\!
   \left( \d m_{\sf_1}^2 \!-\! \d m_{\sf_2}^2 \right)\sin 2\theta_\sf
\\[2mm]
   &&+ \left(  m_{\sf_1}^2 \!-\! m_{\sf_2}^2 \right)
   \cos 2\theta_\sf \, \d\theta_\sf \,.
\end{eqnarray}
In the on--shell scheme the renormalization condition for the 
electroweak gauge boson sector reads
\begin{eqnarray} 
   \frac{\d g}{g} &=& \frac{\d e}{e} \,+\, \frac{1}{\tan^2\theta_W}
   \left( \frac{\d m_W}{m_W} - \frac{\d m_Z}{m_Z} \right)
\end{eqnarray}
with $m_W$ and $m_Z$ fixed as well as the fermion and sfermion masses
as the physical (pole) masses.
For $\tan\b$ we use the condition \cite{pokorski}
Im$\hat\Pi_{A^0 Z^0}(m_A^2) = 0$  which gives the counter term
\begin{eqnarray} 
   \frac{\d \tan\b}{\tan\b} &=& \frac{1}{m_Z \sin 2\b}\, {\rm Im} 
   \Pi_{A^0 Z^0} (m_{A^0}^2).
\end{eqnarray}
The higgsino mass parameter $\mu$ is renormalized in the chargino 
sector \cite{0104109} where it enters in the 22--element of the 
chargino mass matrix $X$, 
\begin{eqnarray} 
   X = \left(
   \begin{array}{cc}
      M & \sqrt2 m_W \sin\b \\ \sqrt2 m_W \cos\b & {\mu}
   \end{array}\right) \hspace{10mm}
   \rightarrow {\d\mu ~=~ (\d X)_{22}} \,,
\end{eqnarray}
and the counter term of the sfermion mixing angle, 
$\d\theta_{\!\sf}$, is fixed such that it cancels the 
anti--hermitian part of the sfermion wave--function corrections 
\cite{guasch, JHEP9905}, 
\begin{eqnarray}
   \delta \theta_{\sf} & = & \frac{1}{4}\, \left(
   \d Z^\sf_{12} - \d Z^\sf_{21}\right)
   = \frac{1}{2\big(m_{\sf_1}^2 \!-\! m_{\sf_{2}}^2\big)}\, {\rm Re}
   \left( \Pi_{12}^\sf(m_{\sf_{2}}^2) + \Pi_{21}^\sf(m_{\sf_{1}}^2)
    \right) \,.
\end{eqnarray}
The one--loop corrected decay width is then given by
\begin{eqnarray}\label{1loopwidth}
\G(A^0 \rightarrow \tilde{f}_i \,\ {\bar{\!\!\tilde{f}}}_{\!\!j})
&=& \frac{N_C^f\, \kappa (m_{A^0}^2, m^2_{\sf_i}, m^2_{\sf_j})}{16 
\,\pi\, m^3_{A^0}}\left[ |G_{ij}^{\sf}|^2 + 2 {\rm Re} \left( 
G_{ij}^{\sf}\cdot \d G_{ij}^{\sf} \right) \right] \,, 
\end{eqnarray}
The infrared divergences in eq.~(\ref{1loopwidth}) are cancelled 
by the inclusion of real photon emission, see the last two Feynman 
diagrams of Fig.~\ref{vertex-graphs}. The decay width of $A^0(p) 
\rightarrow \tilde{f}_i(k_1) +\ {\bar{\!\!\tilde{f}}}_{\!\!j}(k_2) 
+ \g(k_3)$ can be written as
\begin{eqnarray}\non
\G(A^0 \rightarrow \tilde{f}_i \,\ 
{\bar{\!\!\tilde{f}}}_{\!\!j}\,\g) &=& 
\frac{(e\,e_f)^2\,N_C^f\,|G_{ij}^{\sf}|^2}{16\,\pi^3\,m_{A^0}} 
\left[ \left( m_{A^0}^2\!-\!m_{\sf_i}^2\!-\!m_{\sf_j}^2 \right) 
I_{12} \!-\!m_{\sf_i}^2 I_{11}\!-\!m_{\sf_j}^2 
I_{22}\!-\!I_1\!-\!I_2\right] 
\\  
\end{eqnarray}
with the phase--space integrals $I_n$ and $I_{mn}$ defined as 
\cite{Denner} 
\begin{eqnarray}
I_{i_1\ldots i_n}=\frac{1}{\pi^2}
\int\frac{d^3k_1}{2E_1}\frac{d^3k_2}{2E_2}\frac{d^3k_3}{2E_3}
\delta^4(p-k_1-k_2-k_3)\frac{1}
{(2k_3k_{i_1}+\lambda^2)\ldots(2k_3k_{i_n}+\lambda^2)}.
\end{eqnarray}
The corrected (UV-- and IR--convergent) decay width is then given 
by
\begin{eqnarray}
\G^{\rm corr}(A^0 \rightarrow \tilde{f}_i \,\ 
{\bar{\!\!\tilde{f}}}_{\!\!j}) &\equiv& \G(A^0 \rightarrow 
\tilde{f}_i \,\ {\bar{\!\!\tilde{f}}}_{\!\!j}) \,+\, \G(A^0 
\rightarrow \tilde{f}_i \,\ {\bar{\!\!\tilde{f}}}_{\!\!j}\,\g)\,.
\end{eqnarray}

In the following numerical examples, we assume $M_{{\ti Q}_{1,2}} 
= M_{{\ti U}_{1,2}} = M_{{\ti D}_{1,2}} = M_{{\ti L}_{1,2}} = 
M_{{\ti E}_{1,2}}$, $M_{{\ti Q}} \equiv M_{{\ti Q}_{3}} = 
\frac{9}{8} M_{{\ti U}_{3}} = \frac{9}{10} M_{{\ti D}_{3}} = 
M_{{\ti L}_{3}} = M_{{\ti E}_{3}}$ for the first, second and third 
generation soft SUSY breaking masses and $A \equiv A_t = A_b = 
A_\tau$. We take $m_t = 175$ GeV, $m_b = 5$ GeV, $m_Z = 91.2$ GeV, 
$m_W = 80$ GeV and $\sin^2 \theta_W = 0.23$ for Standard Model 
values and the gaugino unification relation $M' = 
{\displaystyle{\frac{5}{3}}} \, \tan^2\theta_W M$.\\ In 

\begin{figure}[h!]
\begin{picture}(160,60)(0,0)
    \put(30,7){\mbox{\resizebox{9cm}{!}
    {\includegraphics{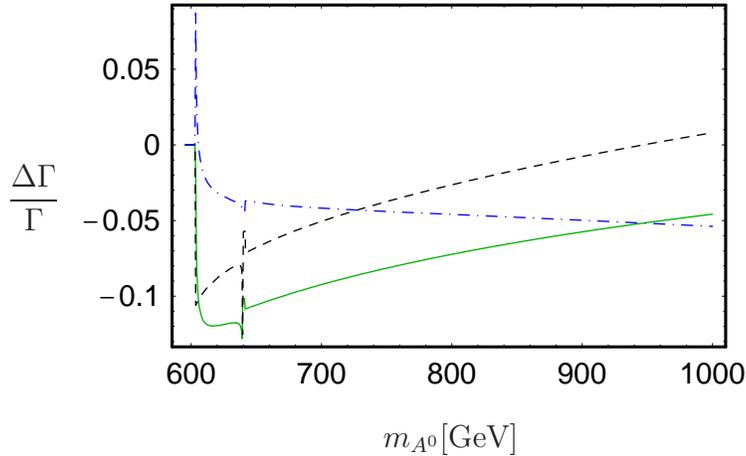}}}}
    \put(80,5){\makebox(0,0)[t]{{$m_{A^0}$[GeV]}}}
    \put(28,35){\makebox(0,0)[r]{{$\displaystyle\frac{\D\Gamma}{\G}$}}}
\end{picture}
\caption{Relative corrections to $A^0 \rightarrow \st_1 
\!\bar{\,\st_2}$, separated into leading Yukawa (black dashed 
line) and the remaining electroweak (blue dash-dotted line) 
corrections. The green solid line corresponds to the full 
electroweak corrections. \label{yukawa_vs_ew}} 
\end{figure}

Fig.~\ref{yukawa_vs_ew} we show the $m_{A^0}$--dependence of the 
relative correction to $A^0 \rightarrow \st_1 \!\bar{\,\st_2}$, 
separated into leading Yukawa and the remaining electroweak 
corrections using $\tan\b = 7$ and $\{ M_{{\ti Q}_1}, M_{\ti Q}, 
A, M, \mu \} = \{1500,300, -500, 120, -260\} \ \mbox{GeV}$ as 
input parameters. As can be seen for larger values of $m_{A^0}$, 
the remaining electroweak corrections can become bigger than the 
leading Yukawa corrections and need to be included. 

\begin{figure}[h!]
\begin{picture}(160,60)(0,0)
    \put(33,5){\mbox{\resizebox{8.5cm}{!}
    {\includegraphics{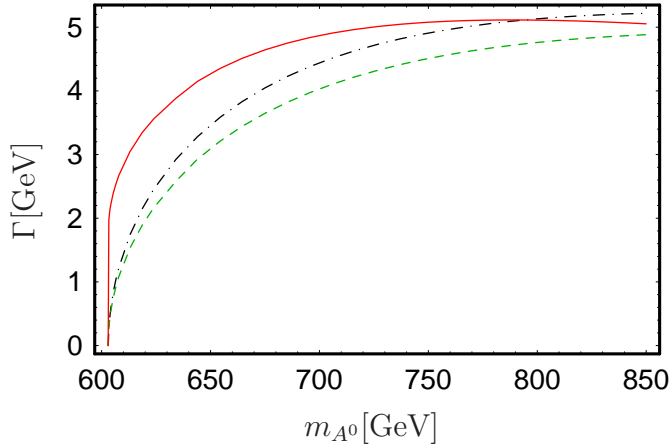}}}}
    \put(76,3){\makebox(0,0)[t]{{$m_{A^0}$[GeV]}}}
    \put(28,26){\rotatebox{90}{{{$\Gamma$[GeV]}}}}
\end{picture}
\caption{Tree--level (black dash-dotted line), full electroweak 
corrected (green dashed line) and full one--loop (electroweak and 
SUSY--QCD) corrected (red solid line) decay width of $A^0 
\rightarrow \st_1 \!\bar{\,\st_2}$. \label{mA_dependence_os}} 
\end{figure}

\noindent In Fig.~\ref{mA_dependence_os}, in addition to the 
tree--level and electroweak corrected decay width for $A^0 
\rightarrow \st_1 \!\bar{\,\st_2}$ we have also included SUSY--QCD 
corrections from \cite{SUSY-QCD}. As input set we have taken the 
same parameters as in Fig.~\ref{yukawa_vs_ew}. Note that the 
electroweak corrections can be of the same size as the QCD 
corrections. 

\begin{figure}[h!]
\begin{picture}(160,60)(0,0)
    \put(32,5){\mbox{\resizebox{8.5cm}{!}
    {\includegraphics{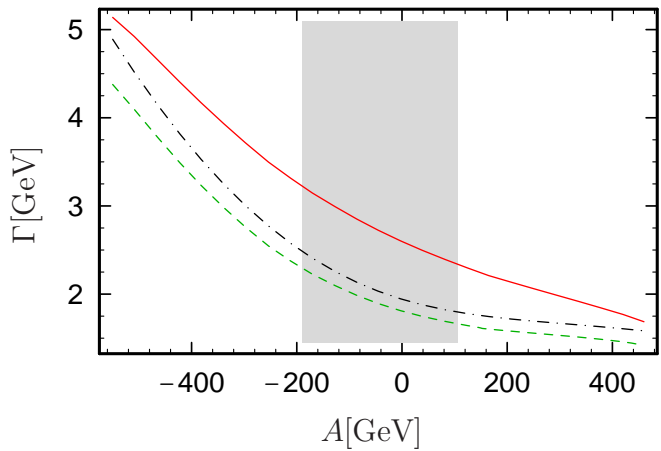}}}}
    \put(76,3){\makebox(0,0)[t]{{$A$[GeV]}}}
    \put(28,26){\rotatebox{90}{{{$\Gamma$[GeV]}}}}
\end{picture}\caption{$A$--dependence of tree--level (black dash-dotted
line), full electroweak corrected (green dashed line) and full 
one--loop (electroweak and SUSY--QCD) corrected (red solid line) 
decay width of $A^0 \rightarrow \st_1 \!\bar{\,\st_2}$. The gray 
area is excluded by phenomenology. \label{A_dependence_os}} 
\end{figure}

\noindent In Fig.~\ref{A_dependence_os} we show the tree--level 
(black dash-dotted line), the full electroweak (green dashed line) 
and the full one--loop corrected (electroweak and SUSY--QCD, red 
solid line) decay width of $A^0 \rightarrow \st_1 \!\bar{\,\st_2}$ 
as a function of $A$. As can be seen electroweak corrections do 
not strongly depend on the parameter $A$ and are almost constant 
about 8\%. As input parameters we have chosen the values given 
above and $m_{A^0} = 700$ GeV.

\section{Conclusions}
\vspace{2mm} In conclusion, we have calculated the full 
electroweak one--loop corrections to {\mbox{$A^0 \rightarrow \st_1 
\!\bar{\,\st_2}$}. We found that in a wide region of parameter 
space electroweak corrections can go beyond $10\%$ and therefore 
have to be included.\\ 

\noindent {\bf Acknowledgements}\\ \noindent The authors 
acknowledge support from EU under the HPRN-CT-2000-00149 network 
programme and the ``Fonds zur F\"orderung der wissenschaftlichen 
Forschung'' of Austria, project No. P13139-PHY.

\end{document}